\documentclass[aps, prb, twocolumn, amsmath,amssymb,showpacs,groupedaddress]{revtex4-1}

\usepackage{graphicx}
\usepackage{color}
\usepackage{amsthm}
\usepackage{amsmath}
\usepackage{amssymb}
\usepackage{setspace}
\usepackage{braket}
\usepackage{float}

\makeatletter
\@ifundefined{textcolor}{}
{
 \definecolor{BLACK}{gray}{0}
 \definecolor{WHITE}{gray}{1}
 \definecolor{RED}{rgb}{1,0,0}
 \definecolor{GREEN}{rgb}{0,1,0}
 \definecolor{BLUE}{rgb}{0,0,1}
  \definecolor{purple}{rgb}{0.5,0,1}
 \definecolor{CYAN}{cmyk}{1,0,0,0}
 \definecolor{MAGENTA}{cmyk}{0,1,0,0}
 \definecolor{YELLOW}{cmyk}{0,0,1,0}
}

\newcommand{\E}{\mathcal{E}}

\newcommand{\bk}{\mathbf{k}}

\newcommand{\kb}{k_{B}}

\begin{document}

\title{The semiclassical theory of anomalous transport\\
in type-II topological Weyl semimetals}

\author{Timothy M. McCormick}
\email[]{mccormick.288@osu.edu}
\thanks{These two authors contributed equally}
\affiliation{Department of Physics and Center for Emergent Materials, The Ohio State University, Columbus, OH 43210, USA}

\author{Robert C. McKay}
\thanks{These two authors contributed equally}
\affiliation{Department of Physics and Center for Emergent Materials, The Ohio State University, Columbus, OH 43210, USA}

\author{Nandini Trivedi}
\email[]{trivedi.15@osu.edu}
\affiliation{Department of Physics and Center for Emergent Materials, The Ohio State University, Columbus, OH 43210, USA}

\date{\today}

\begin{abstract}
Weyl semimetals possess low energy excitations which act as monopoles of Berry curvature in momentum space.  These emergent monopoles are at the heart of the extensive novel transport properties that Weyl semimetals exhibit.  The singular nature of the Berry curvature around the nodal points in Weyl semimetals allows for the possibility of large anomalous transport coefficients in zero applied magnetic field.  Recently a new class, termed type-II Weyl semimetals, has been demonstrated in a variety of materials, where the Weyl nodes are tilted.  We present here a study of anomalous transport in this new class of Weyl semimetals.  We find that the parameter governing the tilt of these type-II Weyl points is intimately related to the zero field transverse transport properties.  We also find that the temperature dependence of the chemical potential plays an important role in determining how the transport coefficients can effectively probe the Berry curvature of the type-II Weyl points.  We also discuss the experimental implications of our work for time-reversal breaking type-II Weyl semimetals.
\end{abstract}
\maketitle

\section{Introduction}

Topological Weyl semimetals have sparked tremendous recent interest in condensed matter physics\cite{felserReview,hasanReview,vishReview}.  These materials host low energy excitations with massless, linear dispersions, known as Weyl fermions\cite{Weyl1929}.  A Weyl node is a monopole of Berry curvature, which acts as a magnetic field in momentum space.  Weyl nodes must come in pairs of opposite chirality\cite{Nielsen} and the sign of their monopole charge corresponds to their chirality.  A direct consequence of the topological nature of the bulk Berry curvature in Weyl semimetals is the presence of topologically protected Fermi arcs that reside in the surface Brillouin zone and form open contours of states\cite{wanTurnVish}.  The bulk Weyl fermions and the surface Fermi arcs provide the key signatures of Weyl semimetals and are responsible for their many novel features.

Weyl fermions have been predicted in a variety of condensed matter systems\cite{wanTurnVish,burkBal,PhysRevLett.107.186806,Volovik2014514,PhysRevX.5.011029,Huang2015} and were first experimentally realized in the transition metal monopnictides, where signatures of the bulk nodes and the surface Fermi arcs were detected by angle-resolved photo emission spectroscopy\cite{Lv2015,Xu2015,Xue1501092,Liu2016}.  Shortly after their discovery, a new class of Weyl semimetals, called type-II Weyl semimetals, was predicted\cite{PhysRevB.78.045415,Soluyanov2015,PhysRevLett.117.056805,Sun2015} and experimentally discovered\cite{Bruno2016,Huang2016,Wu2016,Wang2016} in the transition metal dichalcogenides MoTe$_2$ and WTe$_2$.  Subsequently, a number of other examples of this new class of Weyl semimetals have been discovered\cite{PhysRevLett.117.066402,Xue1603266,PhysRevB.93.201101,Change1600295,ybmnbi2}.  In type-II Weyl semimetals, the Weyl nodes are tilted and therefore have a finite density of electrons and holes at the Weyl energy.

The bulk Berry curvature of Weyl fermions is known to result in a plethora of unique transport phenomena in Weyl semimetals.  In parallel magnetic and electric fields, Weyl semimetals exhibit negative longitudinal magnetoresistance as a result of the chiral anomaly\cite{PhysRevB.89.195137,PhysRevB.88.104412,PhysRevB.92.075205,PhysRevB.89.085126,Zhang2016,PhysRevX.5.031023,Hirschberger2016}.  The surface Fermi arcs also lead to a remarkable mixed real- and momentum-space channel of charge transport\cite{pkv,quantOsc2,qosc3,Moll2016} and have also been predicted to lead to a novel mechanism for entropy transport\cite{arcTherm}.  In bulk thermoelectric transport, Weyl semimetals have been predicted to have a number of distinct signatures\cite{fieteThermoelec,PhysRevB.93.035116,doubleWeyl}, most notably a Nernst thermopower at zero applied magnetic field in time-reversal breaking Weyl semimetals\cite{PhysRevB.93.035116}.  Recent experiments have shown extraordinary thermoelectric properties of NbP, including a large ordinary Nernst effect\cite{nbpNernst} and evidence for a mixed axial-gravitational anomaly\cite{gravAnom}.  

Although there have been some preliminary predictions of transport in type-II Weyl semimetals\cite{Soluyanov2015,PhysRevLett.116.236401,PhysRevB.96.045112,Zyuzin2016}, it remains comparatively less well-understood.  There have been signatures of the chiral anomaly in WTe$_{1.98}$\cite{PhysRevLett.118.096603} as well as evidence of viscous electronic and thermal transport in the type-II Weyl semimetal WP$_2$\cite{planckDis}.   There are also strong candidates for type-II time-reversal breaking Weyl semimetals\cite{ybmnbi2,mn3sndft} and the type-II Weyl semimetal candidates Mn$_3$Sn and Mn$_3$Ge have shown tantalizing signatures of a large anomalous Hall effect\cite{mn3snHall,PhysRevApplied.5.064009,Nayake1501870}.  Furthermore, experimental signatures of the anomalous Nernst effect and anomalous thermal Hall effect have also been detected in Mn$_3$Sn\cite{mn3snnernst}.  

Although ferrmomagnetic metals are known to possess anomalous transport coefficients in zero field\cite{halltop,hallrmp,sarahNernst}, Mn$_3$(Ge,Sn) is instead a weakly canted \textit{antiferromagnet}.  It has been suggested that the real-space magnetic texture can account for a large anomalous Hall effect in Mn$_3$Ge if the spins are non-coplanar\cite{0295-5075-108-6-67001}, however experiments have shown that the Mn$_3$(Ge,Sn) system does possess a large anomalous Hall effect in the planar magnetic phase with a Hall coefficient that is much larger than its weakly canted moment would suggest\cite{mn3snHall}. Thus many puzzles remain.

Motivated in part by some of these experimental puzzles, we study anomalous transport in a lattice model of a time-reversal breaking Weyl semimetal\cite{mkt}.  The model we use allows for tuning through the type-I to type-II transition as well as between different type-II phases with distinct Fermi surface connectivities.  Although the Berry curvature of a Weyl node is independent of its type, the occupation of states immediately surrounding the Weyl nodes is strongly dependent on the precise Fermiology of the material.  Furthermore, the chemical potential in low density systems, such as semimetals, is a strong function of temperature\cite{nbpNernst,wte2liftrans}.  A detailed understanding of the interplay between the tilt of the nodes, the connectivity of the Fermi pockets, and the temperature dependence of the chemical potential is required for a complete theoretical picture of anomalous transport in type-II Weyl semimetals.

\begin{figure*}
	\centering
	\includegraphics[width=1.0\textwidth]
	{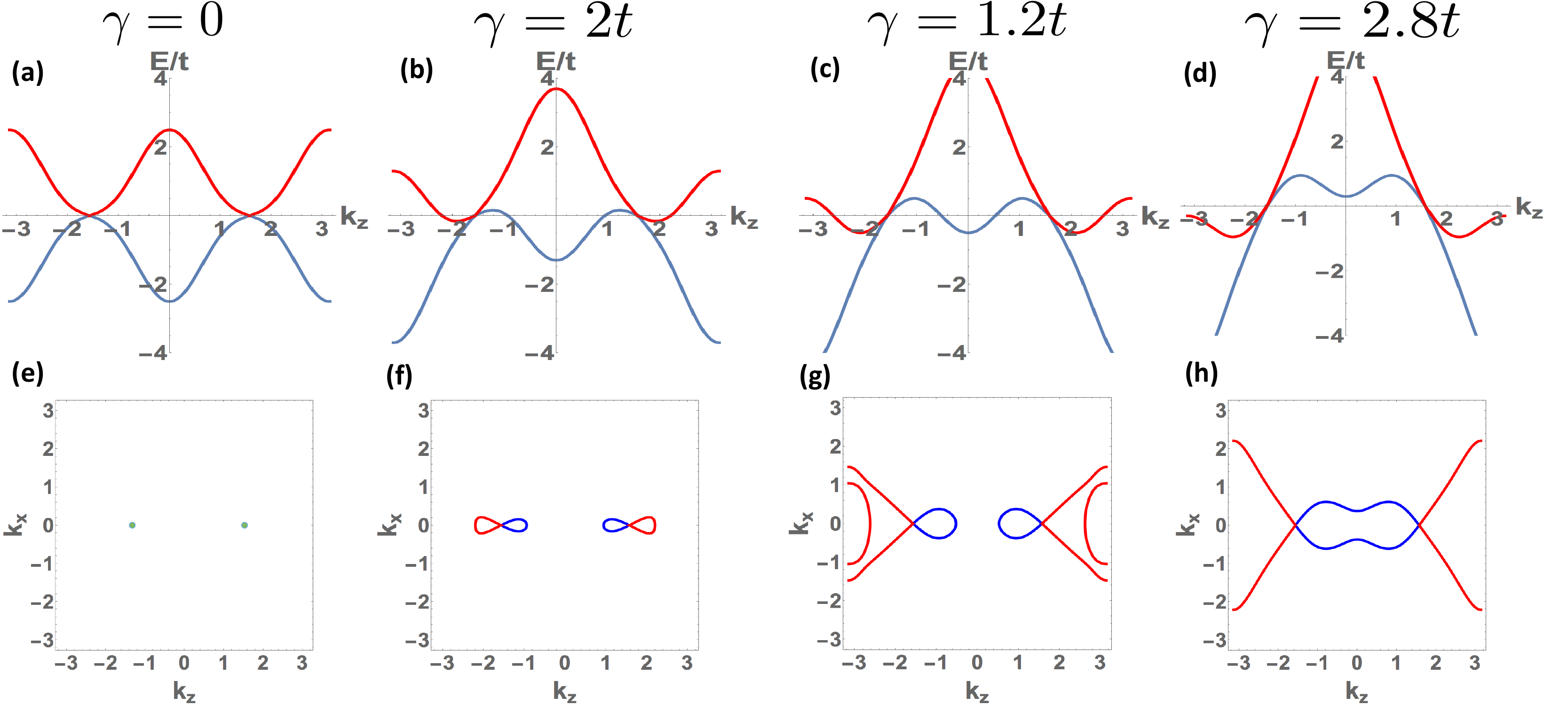}
	\caption{Cuts through the band structure given by the Hamiltonian in Eqn. (\ref{ham}).  In (a-d), we show energy versus $k_z$ cuts for $k_x = k_y = 0$.  Here we have chosen $m = 3t$; $t_z = t$; $k_0 a = \pi/2$; $\gamma_z = 0.5t$ and $\gamma = 0$ (a), $\gamma = 1.2t$ (b), $\gamma = 2t$ (c), and $\gamma = 2.8t$ (d).  In (e-h), we show constant energy cuts for the band structure defined by Eqn. (\ref{ham}).  (a) and (e) are in the type-I limit; (b) and (f) are in the type-II limit with distinct pockets making up each nodes; (c) and (g) are in the type-II limit after the electron pockets have merged; and (d) and (h) are in the type-II regime where the Weyl nodes share only a single electron and single hole pocket. Thus, as $\gamma$ is increased we can see the successive Lifshitz transitions described in the text.}
	\label{bandStructures}
\end{figure*}

The remainder of this paper is organized as follows.  In section II we present a time-reversal breaking model for a type-II Weyl semimetal that allows for tuning between various Fermi pocket connectivities.  We discuss how the chemical potential depends on temperature for this model.  In section III, we investigate how the topological features determine anomalous transverse transport coefficients in the type-I and type-II regimes, including the cross-over between them.  For the model described in section II, we calculate the anomalous Hall coefficient, the anomalous transverse thermoelectric coefficient, and the thermal Hall coefficient.  In section IV, we discuss our results and conclude with experimental implications.

\section{Model}

In the continuum limit, a type-II Weyl node can be described by the following Hamiltonian
\begin{equation}
\label{contham}
\hat{H}_W = \gamma k_z \hat{\sigma}_0 +\chi \hbar v_F \big(\bk - \mathbf{k}_W \big) \cdot \mathbf{\sigma}, 
\end{equation}
where $\sigma_0$ is the $2\times2$ identity matrix, $\mathbf{\sigma}$ is the vector of Pauli matrices, $\gamma$ describes the tilt of the node, $\mathbf{k}_W$ is the momentum of the node, $v_F$ is the Fermi velocity, taken to be a constant, and $\chi$ is the chirality of the node.  The type-I to type-II transition for this continuum model occurs at $\gamma = \hbar v_F$.  For $\gamma = 0$, the continuum model above has proven to be profoundly useful in understanding the properties of type-I Weyl semimetals.  However, in the type-II case, this continuum model is manifestly unphysical.  For $\gamma > \hbar v_F$, Eqn. (\ref{contham}) describes an electron and a hole pocket that are unbounded and never close at large $\bk$.  Since various aspects of type-II Weyl semimetals are strongly dependent on the nature of the extended Fermi pockets surrounding the nodes\cite{mkt}, it is necessary to instead consider a lattice model.

In order to avoid the difficulties noted above caused by the unbounded Fermi pockets of continuum models for type-II Weyl semimetals, we consider the following model for a time-reversal symmetry breaking system with two Weyl nodes:
\begin{multline}
\label{ham}
\hat{H} = \gamma(\cos(k_z a) - \cos(k_0 a)) \hat{\sigma}_{0} - 2t \sin(k_x a) \hat{\sigma}_{1} \\- 2t \sin(k_y a) \hat{\sigma}_{2}  - \Big[ 2 t_{z} \big( \cos(k_{z} a) - \cos(k_{0} a) \big) \\+ m (2 - \cos(k_{x} a) - \cos(k_y a))  + \gamma_z (\cos(3 k_{z} a) - \cos(3 k_{0} a)) \Big]\hat{\sigma}_{3}
\end{multline}
where $\sigma_0$ is the $2\times2$ identity matrix, $\hat{\sigma}_j$ is the $j$-th Pauli matrix, $a$ is the lattice spacing, $t$, $m$, $\gamma_z$ and $t_z$ are hopping amplitudes, $k_0$ sets the node separation, and $\gamma$ sets the tilt of the Weyl nodes. This model supports two sets of electron and hole pockets, with each Weyl point being comprised of a separate pair of electron and hole pockets. The type-I to type-II Lifshitz transition happens at $\gamma = 2t_z - 3\gamma_z$. 

\subsection{Lifshitz transitions}

In Fig. \ref{bandStructures}, we plot the energy band structure given by Eqn. (\ref{ham}) for the parameters $m = 3t$, $t_z = t$, $k_0 a = \pi/2$, and $\gamma_z = 0.5t$ for several values of $\gamma$.  For these model parameters, the type-I to type-II transition occurs at $\gamma = 0.5t$.  There is also a second set of topological Lifshitz transitions where the separate electron and hole pockets each merge into a single pocket.  A hole pocket within the electron pocket also forms for a small range of parameters.  For the model parameters defined above, the elecron pockets merge at $\gamma \approx 2t$ and the hole pockets similarly merge into a single Fermi pocket at $\gamma \approx 2.5t$.  The tunability of this model between these four different regimes makes it the minimal model to understand type-II Weyl semimetals.  We will show how each regime has its own signature in anomalous transport. 

\subsection{Temperature dependence of the chemical potential}

In metals with high densities of electrons at degenerate temperatures, the chemical potential is nearly constant with respect to temperature.  However, in low density semimetals, it has been shown that the chemical potential has a strong temperature dependence at experimentally relevant temperatures\cite{nbpNernst,wte2liftrans}.  For the lattice model in Eqn. (\ref{ham}), we calculate the temperature dependence of the chemical potential by self-consistently solving for $\mu (T)$ for fixed density:
\begin{equation}
\label{muoft}
n = \int_{-\infty}^{\infty} d\E \dfrac{g(\E)}{1+e^{\frac{\E-\mu(T)}{\kb T}}},
\end{equation}
where $n$ is the density, $T$ is the temperature and $g(\E)$ is the density of states found through
\begin{equation}
\label{doseqn}
g(\E) = -\dfrac{1}{\pi} \sum_n \textrm{Im}\Bigg[
\int \dfrac{d^3 k}{(2\pi)^3} G_{n}(\bk,\E)
\Bigg],
\end{equation}
where $G_{n}(\bk,\E)$ is the Green function of the $n$-th band.

\begin{figure*}
	\centering
	\includegraphics[width=1\textwidth]
	{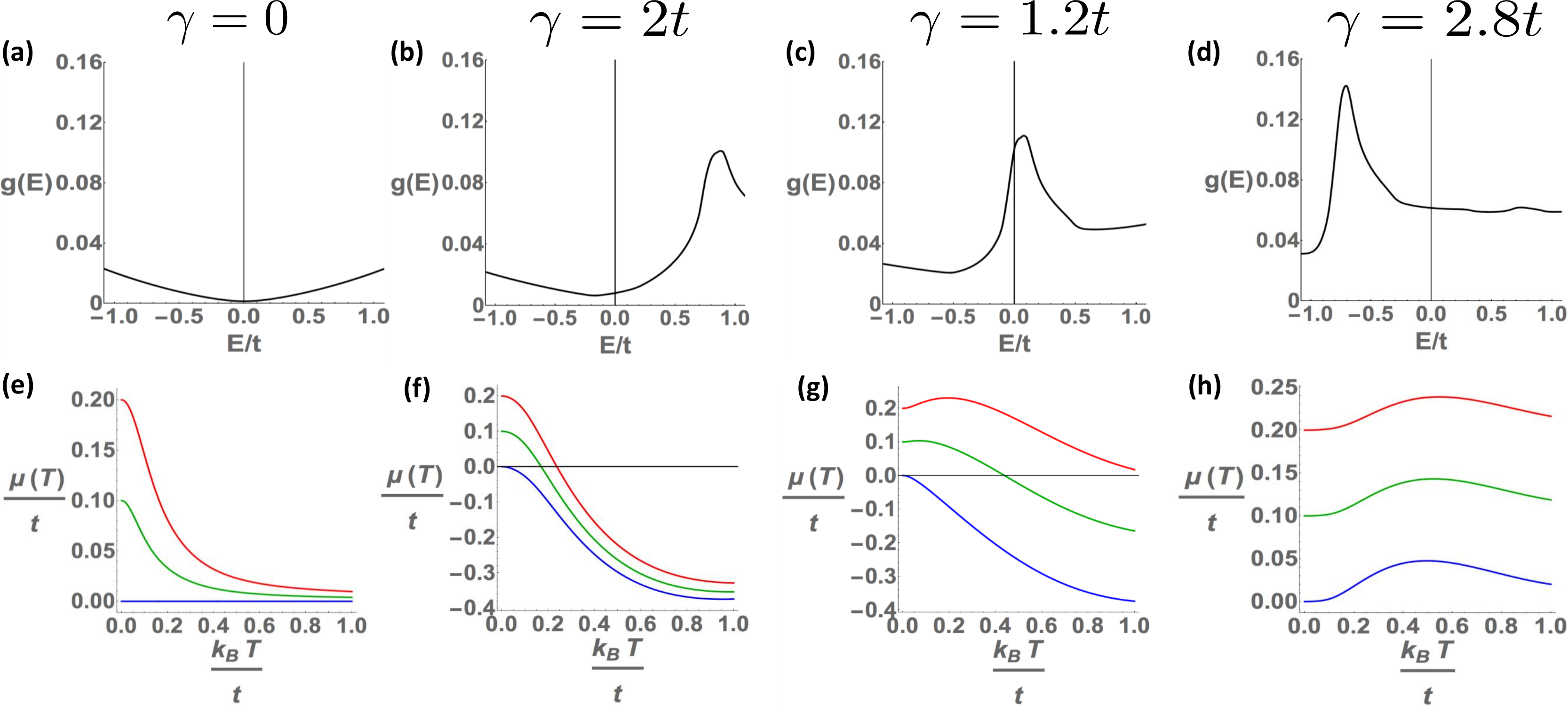}
	\caption{The density of states for $m = 3t$, $t_z = t$, and $k_0 a = \pi/2$, and $\gamma_z = 0.5t$ is shown in (a-d) for different values of the tilt parameter: $\gamma = 0$ (a), $\gamma = 1.2t$ (b), $\gamma = 2t$ (c), and $\gamma = 2.8t$ (d).  (e-h) illustrates the temperature dependence of the chemical potential, $\mu$(T), for the same values of $\gamma$ as (a-d).  Each plot shows three separate values of $E_F$: $E_F=0$ (blue), $E_F=0.1$ (green), and $E_F=0.2$ (red).  We see that for smaller values of $\gamma$, $g(\E)$ has a minimum close to the Weyl energy $\E = 0$, but for larger values of $\gamma$, this minimum shifts far from the nodes.  This has a strong effect on the shift of the chemical potential with temperature.}
	\label{mu_and_dos}
\end{figure*}

For isolated type-I Weyl points, the minimum of $g(\E)$ will generically occur at the Weyl points\cite{nbpNernst} due to the symmetry of the particle and hole bands. This results in the chemical potential shifting to the Weyl points with increasing temperature in type-I Weyl semimetals.  However, in general, the hole and electron pockets are not symmetric about the Weyl energy in type-II Weyl semimetals, and the minimum of $g(\E)$ will occur above or below the Weyl nodes.  In Fig. \ref{mu_and_dos}a-d we plot the density of states $g(\E)$ of the model in Eqn. (\ref{ham}) for several values of the tilt parameter $\gamma$ around the node energy.  For $\gamma = 0$, we see the density of states is minimum at the Weyl energy.  However, for larger values of $\gamma$, in the type-II regime, the tilt of the Weyl cones breaks particle hole symmetry and shifts the minimum of $g(\E)$ away from the Weyl energy. In the large $\gamma$ limit where both the electron and hole pockets merge (Fig. \ref{mu_and_dos}d), the minimum of the density of states shifts far from the Weyl energy.  In Fig. \ref{mu_and_dos}e-h we show how the chemical potential evolves with temperature for each value of $\gamma$.  For low values of $\gamma$ in Fig. \ref{mu_and_dos}e and \ref{mu_and_dos}f, we see that the chemical potential shifts to the Weyl energy roughly on a temperature scale of the distance $E_F$ is away from the energy for which $g(\E)$ is minimized.  However, as the nodes become increasingly tilted, the temperature scale over which $\mu(T)$ shifts becomes much larger than the relevant scales in transport that we will consider.  

\subsection{Berry curvature}

We can rewrite the Hamiltonian in Eqn. (\ref{ham}) as $\hat{H} = d_{0}(\bk) \hat{\sigma}_{0} + \mathbf{d}(\bk) \cdot \mathbf{\sigma}$ and upon doing so we can calculate the Berry curvature of the $n$-th band through\cite{bernBook}
\begin{equation}
\label{berryDef}
\Omega_{n,i}(\bk) = \epsilon_{ijl} (-1)^{n} \dfrac{\mathbf{d}\cdot \left( \partial_{k_j} \mathbf{d} \times \partial_{k_l} \mathbf{d} \right)}{2|\mathbf{d}|^3},
\end{equation}
where $\epsilon_{ijl}$ is the perfectly antisymmetric tensor.
The Berry curvature plays a central role in the theory of anomalous transport.  The parameter $\gamma$ which determines the tilt, and therefore the type of the Weyl nodes, is embedded in $d_0(\bk)$, which does not enter Eqn. (\ref{berryDef}).  We note that due to the extended nature of the pockets in type-II Weyl semimetals, the anomalous transport coefficients are strongly dependent on the details of the Fermiology. 

\subsection{Definition of transport coefficients}

From Onsager's generalized transport equations\cite{harmHonig}, we have
\begin{equation}
\label{onsagerDef}
\left( \begin{array}{c}
\mathbf{J}^e \\
\mathbf{J}^Q 
\end{array} \right)
=
\left( \begin{array}{cc}
\mathbf{L}^{EE} & \mathbf{L}^{ET} \\
\mathbf{L}^{TE} & \mathbf{L}^{TT}
\end{array} \right)
\cdot 
\left( \begin{array}{c}
\mathbf{E} \\
-\nabla T 
\end{array} \right),
\end{equation}
where $\mathbf{J}^e$ is the charge current density, $\mathbf{J}^Q$ is the heat current density, $\mathbf{E}$ is the electric field and $\nabla T $ is the applied temperature gradient.  In general, the transport coefficients in Eqn. (\ref{onsagerDef}) must be obtained by solving for the non-equilibrium distribution function using the Boltzmann formalism.  However, in the absence of an applied magnetic field, the situation simplifies dramatically.  In this case, the modified velocity of the $n$-th band is found to be 
\begin{equation}
\mathbf{\dot{r}}_{n} = \mathbf{v}_{n}(\bk) + \dfrac{e}{\hbar} \big(\mathbf{E} \times \mathbf{\Omega}_{n}(\bk) \big),
\end{equation}
where $\mathbf{v}_{n}(\bk) \equiv \frac{1}{\hbar}\mathbf{\nabla}_\bk \E_{n}(\bk)$ is the usual group velocity and where $\dfrac{e}{\hbar} \big(\mathbf{E} \times \mathbf{\Omega}_{n}(\bk) \big)$ is the anomalous velocity due to the Berry curvature\cite{berryrmp}. The Berry curvature is sufficient to generate nonzero transverse transport coefficients $L^{EE}_{xy}$, $L^{ET}_{xy}$, and $L^{TT}_{xy}$ in the presence of zero external magnetic field.  We consider the effects in type-II Weyl semimetals on each of these coefficients below.  For the remainder of this paper, we will only consider the case of zero magnetic field.

\section{Anomalous Transport}

In this section, we consider anomalous transport in a two band lattice model for the type-II time reversal breaking Weyl semimetal described in Section II.  We calculate the anomalous Hall coefficient $L^{EE}_{xy}$, the anomalous transverse thermoelectric coefficient $L^{ET}_{xy}$, and the anomalous thermal Hall coefficient $L^{TT}_{xy}$.

\subsection{Anomalous Hall Effect}

The anomalous Hall effect has long been studied in the context of Weyl semimetals.  For a time-reversal breaking Weyl semimetal, it was shown that the anomalous Hall conductivity $L^{EE}_{xy}$ was directly proportional to the net separation between Weyl nodes\cite{ranLu2011}.  In the presence of Berry curvature, we have seen that the equations of motion are modified and as a result\cite{KL1954,lutt1958,berryrmp,hallrmp}, the anomalous Hall coefficient takes the following form 
\begin{equation}
\label{leexy}
L^{EE}_{xy} = \frac{e^2}{\hbar} \sum_n \int \dfrac{d^3 k}{(2\pi)^3} \Omega_{n,z}(\bk) f_{0}(\bk),
\end{equation}
where $f_{0}(\bk) = \left(1+e^{\frac{\E(\bk)-\mu}{\kb T}} \right)^{-1}$ is the equilibrium Fermi-Dirac distribution. 

\begin{figure*}
	\centering
	\includegraphics[width=1.0\textwidth]
	{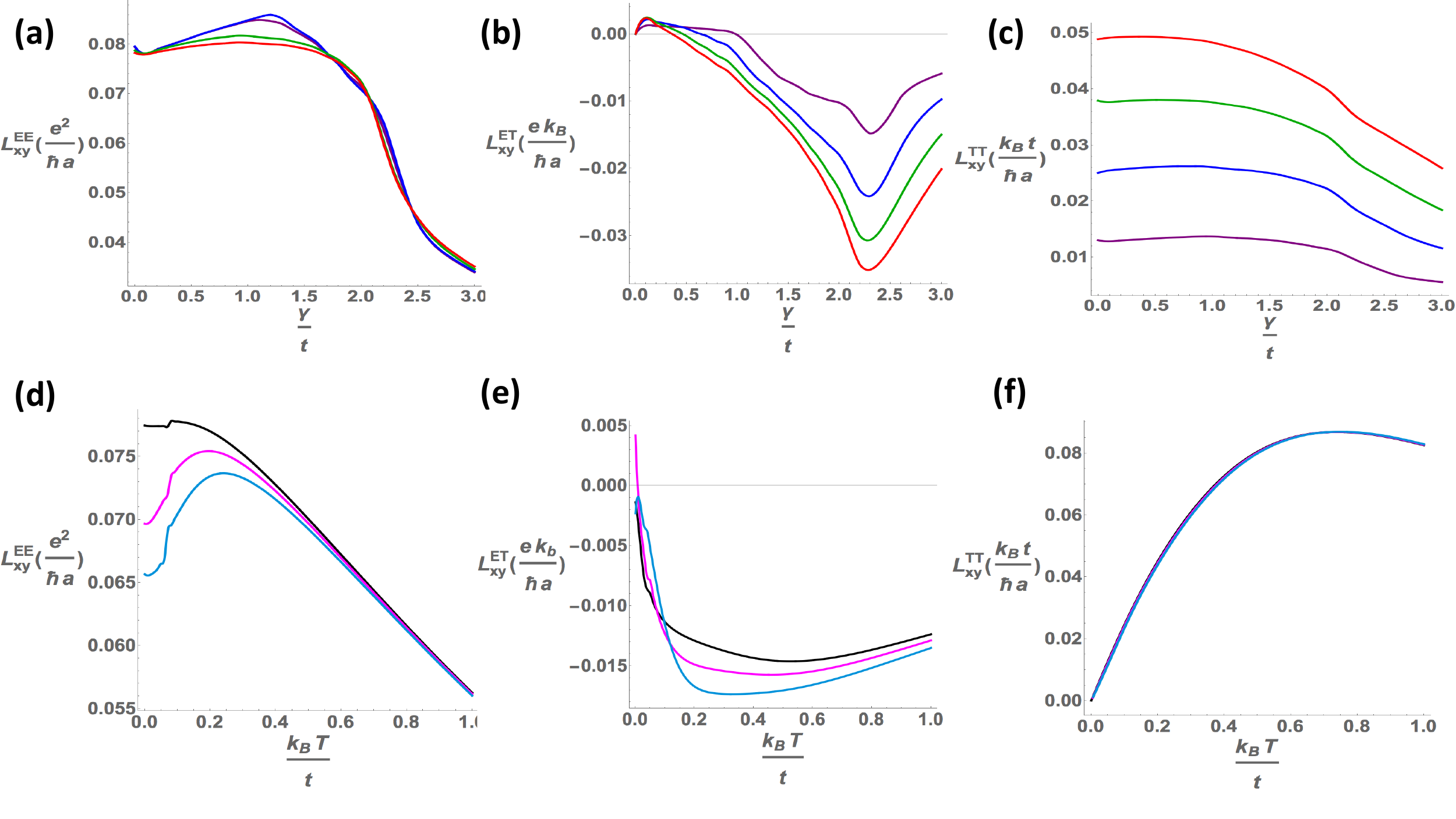}
	\caption{ In (a-c), we plot each anomalous transport coefficient $L^{EE}_{xy}$, $L^{ET}_{xy}$, and $L^{TT}_{xy}$ for the Hamiltonian given by Eqn. (\ref{ham}) with parameters $m = 3t$, $t_z = t$, and $k_0 a = \pi/2$, and $\gamma_z = 0.5t$, as a function of $\gamma$ for the following temperatures $T=0.05 t$ (purple), $T=0.1 t$ (blue), $T=0.15 t$(green), and $T=0.2 t$ (red).  In (d-f), we show $L^{EE}_{xy}$, $L^{ET}_{xy}$, and $L^{TT}_{xy}$ for the same values as in (a)-(c), with $\gamma = 1.2t$, plotted as functions of temperature for various Fermi energies $E_F$: $E_F=0$ (black), $E_F=0.1 t$ (magenta), and $E_F=0.2 t$ (blue).}
	\label{anom_coeff_vs_gamma_T_changes}
\end{figure*}

The anomalous Hall coefficient in Eqn. (\ref{leexy}) integrates the Berry curvature over all filled states.  In the un-tilted regime ($\gamma = 0$), Eqn. (\ref{ham}) will have particle-hole symmetry and since, from Eqn. (\ref{berryDef}), the Berry curvature is opposite in sign but equal in magnitude at a particular point $\bk$ for two bands.  Therefore, equal energy Fermi surfaces at positive and negative energies will have opposite net integrated Berry curvature $\Omega_{n,z}(\bk)$.  Therefore, at $\gamma = 0$ and $E_F = 0$, the filled states all contribute to one sign of $L^{EE}_{xy}$.  As the tilt $\gamma$ is increased, the Fermi surfaces become increasingly asymmetric, and the distributions of $\Omega_{n,z}(\bk)$ over the occupied states changes dramatically.  

In Fig. \ref{anom_coeff_vs_gamma_T_changes}a, we plot the anomalous Hall coefficient $L^{EE}_{xy}$ as a function of the tilt parameter $\gamma$ for several fixed temperatures at $E_F = 0$.  Other than a small increase in $L^{EE}_{xy}$ for $\gamma < 1.5t$ for small temperatures, we note that the curves are substantially similar.  We can understand the increase of $L^{EE}_{xy}$ by noting that the asymmetry in the Berry curvature distribution enhances the Hall coefficient for small temperatures but that this effect is washed out as the temperature increases and the Fermi function broadens.  When the separate electron pockets comprising the Weyl nodes merge into a single electron pocket at $\gamma = 2t$, we see that the Hall coefficient uniformly decreases for all temperatures.   

In Fig. \ref{anom_coeff_vs_gamma_T_changes}d, we show the temperature dependence of the anomalous Hall coefficient $L^{EE}_{xy}$ as a function of temperature for fixed $\gamma = 1.2t$ for various Fermi energies $E_F$.  We see that for $E_F = 0$, $L^{EE}_{xy}$ decreases monotonically as a function of temperature.  This is because as the temperature is raised, the Fermi function broadens and electronic states with the opposite sign of $\Omega_{n,z}(\bk)$ begin to become occupied.  This is true for the type-I case as well.  However, for $E_F \neq 0$, the Hall coefficient will in general be lower than its $E_F = 0$ value at $T \approx 0$.  For $E_F > 0$ in Fig. \ref{anom_coeff_vs_gamma_T_changes}d, we see that $L^{EE}_{xy}$ attains a maximum at $T > 0$.  This is due to the movement of the chemical potential with temperature (see Fig. \ref{mu_and_dos}e-h).  As the chemical potential crosses the Weyl energy, a maximum is attained in the anomalous Hall coefficient as a function of temperature.  This effect is seen to a lesser extent in the type-I case, however there are two distinct differences in the type-II regime: 

\noindent (i) In the type-I case, $\mu(T)$ always shifts to the node energy.  However, in the type-II case, the minimum of the density of states $g(\E)$ occurs generically at a different energy, causing $\mu(T)$ to cross the Weyl energy rather than approach it asymptotically. This leads to a sharper rise of $L^{EE}_{xy}$ with temperature.

\noindent (ii) Due to the higher density of states around the nodes, where the Berry curvature is stronger, the anomalous Hall coefficient is more sensitive to the shift of the chemical potential.  

At higher temperatures, the chemical potential reaches the energy where the density of states is minimized and the temperature dependence of $L^{EE}_{xy}$ is uniform across each value of $E_F$.  Although here we have only shown the results for a single value of $\gamma$, the broad conclusions that we have drawn remain true.  There is a peak in temperature where $\mu(T)$ crosses the node energy.  In the type-II regime, this increase over the $T = 0$ value is, in general, greater than that in the type-I regime for similar other model parameters.  However, we note that for values of $\gamma$ such that the electron (or the hole pockets) have merged, the temperature dependence becomes much weaker.

\subsection{Anomalous Thermoelectric Effect}

In the presence of Berry curvature, the electronic wavefunction acquires an orbital magnetization that is responsible for nontrivial anomalous thermoelectric properties \cite{anomNernstNiu}.
It is found that in this case, the anomalous transverse thermoelectric coefficient is given by 
\begin{equation}
\label{letxy}
L^{ET}_{xy} = \frac{\kb e}{\hbar} \sum_n \int \dfrac{d^3 k}{(2\pi)^3} \Omega_{n,z}(\bk) s(\bk),
\end{equation}
where $s(\bk)$ is the electronic entropy given by
\begin{equation}
\label{entropy}
s(\bk) = -f_0(\bk) \ln(f_{0}(\bk)) - (1 - f_{0}(\bk))\ln(1 - f_{0}(\bk)),
\end{equation}
and where $f_0(\bk)$ is the equilibrium Fermi-Dirac distribution defined above.  The entropy, as defined in Eqn. (\ref{entropy}), is sharply peaked around $\mu(T)$. Therefore, unlike the anomalous Hall coefficient, the anomalous thermoelectric coefficient $L^{ET}_{xy}$ is only sensitive to the Berry curvature around the chemical potential.  

The tilt $\gamma$ of the model for a type-II Weyl semimetal can lead to a change in the distribution of Berry curvature $\Omega_{n,z}(\bk)$ over the occupied states that is drastic enough to flip the  sign of $\Omega_{n,z}(\bk)$ above or below the Weyl node in energy, for energies small compared with the bandwidth $\E \ll t$.  Hence, at energies just above and below the Weyl energy, $\Omega_{n,z}(\bk)$ has the same sign for a tilted Weyl cone.  This leads to a large enhancement of the anomalous thermoelectric coefficient for energies around the node.  We see this in Fig. \ref{anom_coeff_vs_gamma_T_changes}b, where $L^{ET}_{xy}$ is plotted as a function of $\gamma$ for various temperatures.  In the type-I case, $L^{ET}_{xy}$ is small in magnitude and positive in sign.  After the Lifshitz transition to the type-II regime at $\gamma = 0.5t$, we see a change in sign of $L^{ET}_{xy}$ and substantial increase in its magnitude.  This is precisely because, for energies close to the Weyl energy, the integrated net Berry curvature in the $z$-direction $\Omega_{n,z}(\bk)$ has the same sign and a large value.  The anomalous thermoelectric coefficient $L^{ET}_{xy}$ attains a maximum at $\gamma \approx 2.3t$ after the electron pockets have merged and just as the hole pockets are about to merge (with increasing $\gamma$).  Hence, we see that measurable quantities, such as the Nernst effect, which depend on $L^{ET}_{xy}$ may be quite sensitive to Lifshitz transitions between various regimes of Fermi pocket connectivity in type-II Weyl semimetals.

In Fig. \ref{anom_coeff_vs_gamma_T_changes}e, we plot the anomalous thermoelectric coefficient $L^{ET}_{xy}$ as a function of temperature for various Fermi energies $E_F$ at $\gamma = 1.2t$.  Here we see that with increasing temperature there is an increase with temperature in the magnitude of $L^{ET}_{xy}$ for small $T$.  This occurs because, as the entropy $s(\bk)$ broadens, $L^{ET}_{xy}$ is enhanced from the entropy including a larger range of energies in the integral in Eqn. (\ref{letxy}).  At a temperature on the order of the energy for which the density of states $g(\E)$ attains its minimum, $L^{ET}_{xy}$ reaches its maximum absolute value.  For Fermi energies $E_F$ farther from the node energy, we see that the maximum absolute value of $L^{ET}_{xy}$ increases.  This occurs for the same reason as in the anomalous Hall coefficient above; for a type-II Weyl semimetal the chemical potential $\mu(T)$ will pass over the Weyl energy where $\Omega_{n,z}(\bk)$ is enhanced.  

Finally, we also note that since the Berry curvature is symmetric in magnitude but opposite in sign for an un-tilted type-I Weyl node, at exactly $E_F = 0$, $L^{ET}_{xy}$ must vanish.  Hence, at precisely $E_F = 0$, $L^{ET}_{xy}$ can only take a nonzero value for a type-II Weyl semimetal where the shifted occupation of states can lead to a nonzero net Berry curvature when Eqn. (\ref{letxy}) is evaluated at the node energy.

\subsection{Anomalous Thermal Hall Effect}

As in the anomalous transverse thermoelectric effect above, similar effects of orbital magnetization in the presence of Berry curvature also lead to an anomalous thermal Hall effect in the absence of magnetic field\cite{thermhall1,thermhall2,thermhall3}.
In the presence of Berry curvature, the anomalous thermal Hall coefficient is given by
\begin{equation}
\begin{split}
\label{lttxy}
L^{TT}_{xy} = \dfrac{\kb^2 T}{\hbar} \sum_{n} \int \dfrac{d^3 k}{(2\pi)^3} \Omega_{n,z}(\bk) \Bigg(
\dfrac{\pi^2}{3} + \dfrac{(\E - \mu)^2}{(\kb T)^2}f_0(\bk)\\-\textrm{ln}\bigg(1+e^{-\frac{\E-\mu(T)}{\kb T}}\bigg)+2\textrm{Li}_{2}\big( 1 - f_0(\bk) \big)   \Bigg),
\end{split}
\end{equation}
where $f_0(\bk)$ is again the equilibrium Fermi-Dirac distribution and where $\textrm{Li}_{m}(z)$ is the polylogarithm function of order $m$ defined by
\begin{equation}
\textrm{Li}_{m}(z) = \sum_{k=1}^\infty \dfrac{z^k}{k^m}.
\end{equation}

Like the anomalous Hall coefficient, the anomalous thermal Hall coefficient defined by Eqn. (\ref{lttxy}) integrates over many states below the chemical potential.  However, unlike the simple Fermi distribution $f_0(\bk)$, the kernel of the integrand multiplying $\Omega_{n,z}(\bk)$ in Eqn. (\ref{lttxy}) has a broader inflection point at $\E = \mu(T)$ than $f_0(\bk)$ and hence, $L^{TT}_{xy}$ is sensitive to Berry curvature over a wider range than $L^{EE}_{xy}$.  In Fig. \ref{anom_coeff_vs_gamma_T_changes}c, we plot $L^{TT}_{xy}$ as a function of $\gamma$ and we see that it is quite similar with the anomalous Hall coefficient plotted in Fig. \ref{anom_coeff_vs_gamma_T_changes}a.  However, we note that the anomalous thermal Hall coefficient increases with increasing $\gamma$ for all temperatures up to the Lifshitz transition where the electron pockets merge.  We also see that after the Fermi pockets do merge, the decrease of $L^{TT}_{xy}$ is less rapid than $L^{EE}_{xy}$ with increasing $\gamma$.  We also note that the temperature dependence of $L^{TT}_{xy}$ is increasing rather than decreasing as in $L^{EE}_{xy}$ at large $T$. Finally, the dependence of $L^{TT}_{xy}$ on $E_F$ is essentially negligible.

\section{Discussion and Conclusion}

We have calculated how the anomalous transport coefficients reveal signatures of type-II Weyl semimetals.  In particular, the anomalous transverse thermoelectric coefficient $L^{ET}_{xy}$ is greatly enhanced by tilting the Weyl nodes.  However, we have also seen that even deep in the type-II phase, the various anomalous coefficients are sensitive to changes in Fermi surface topology even quite far from the Weyl nodes.  Each of $L^{EE}_{xy}$  and $L^{TT}_{xy}$ show a marked decrease in magnitude as the electron pockets merge, while $L^{ET}_{xy}$ peaks just after they merge and $L^{ET}_{xy}$ decreases for $\gamma$ large enough that the hole pockets similarly merge.

Previous calculations have shown\cite{mkt} that tuning through the various Lifshitz transitions within the type-II Weyl semimetal phase is associated with changing connectivities of the topological Fermi arcs.  The bulk Fermi surface Berry curvature is deeply linked to the topology of the bulk band structure, and it is enlightening to see the anomalous transport coefficients reflect this bulk-boundary correspondence.  In some inversion-breaking Weyl semimetals\cite{Change1600295}, it has been shown that strain may be possible to tune between a type-I to type-II phase transition. Although this has not yet been demonstrated in a time-reversal breaking Weyl semimetal, it should be possible in principle.  This would allow for an experimental verification of our results as well as allowing for deeper explorations of connections between anomalous transport and changes in Fermi arc topology.

We have also seen that the temperature dependence of the anomalous transport coefficients is quite distinct for each of $L^{EE}_{xy}$, $L^{ET}_{xy}$, and $L^{TT}_{xy}$.   The temperature dependence of the chemical potential has already been demonstrated to lie at the heart of the ordinary Nernst thermopower in the Weyl semimetal NbP\cite{nbpNernst}.  However, similar effects in zero field anomalous transport have remained, until this point, unexplored.  We have demonstrated that, through the location of the Fermi energy $E_F$, the temperature at which $L^{EE}_{xy}$ attains its maximum can be tuned, as can the strength of the anomalous thermoelectric coefficient $L^{ET}_{xy}$.  

The generation of some kind of transverse response to an applied gradient in electric potential or thermal gradient unifies the zoo of the various Hall effects.  The ability to generate such a response in the absence of an externally applied magnetic field opens the door to a wide variety of technological applications.  Weyl semimetals have been predicted to generate various anomalous transport phenomena in zero field due to their Berry curvature.  However, the lack of experimental realizations of time-reversal breaking Weyl semimetals have stymied their application.  Recently, however, it has been proposed that time-reversal breaking Weyl semimetal candidates are type-II in nature.  Our calculations serve not only to guide future experiments but also demonstrate that through changing various experimentally accessible properties, it may be possible to tune the anomalous transport coefficients in type-II weyl semimetal to an extent not possible in the type-I case.

\section*{Acknowledgements}  We are grateful to S. J. Watzman and J. P. Heremans for very stimulating discussions.  T. M. M. was supported by the Center for Emergent Materials, an NSF MRSEC, under grant DMR-1420451.  R. C. M. acknowledges funding from the OSU Robert P. Caren and Family scholarship.  N. T. acknowledges funding from NSF-DMR-1309461.

\bibliography{anomTransportTypeII_ref}

\end{document}